%% file: main.tex
\documentclass[conference]{IEEEtran}

\usepackage{amsmath,amssymb,epsfig,psfrag,cite,subfigure}
\include{macros}
\usepackage{graphicx}
\usepackage{epstopdf}

\usepackage{bm}

\usepackage{geometry}
\usepackage{tcolorbox}

\usepackage{soul}

\usepackage{amsthm}

\usepackage{nicefrac}

\usepackage{lipsum,multicol}

\usepackage{algorithm}
\usepackage{algpseudocode}

\algdef{SE}[SUBALG]{Indent}{EndIndent}{}{\algorithmicend\ }%
\algtext*{Indent}
\algtext*{EndIndent}

\usepackage{mwe}    
\usepackage{epsfig,psfrag}
\usepackage{subfigure}
\usepackage{color}
\usepackage{url}
\usepackage{mathtools,xparse}

\usepackage{gensymb}

\usepackage[multiple]{footmisc}

\usepackage{xr}

\makeatletter
\newcommand*{\addFileDependency}[1]{
  \typeout{(#1)}
  \@addtofilelist{#1}
  \IfFileExists{#1}{}{\typeout{No file #1.}}
}
\makeatother

\newcommand*{\myexternaldocument}[1]{%
    \externaldocument{#1}%
    \addFileDependency{#1.tex}%
    \addFileDependency{#1.aux}%
}

\myexternaldocument{supp}

\usepackage{accents}

\makeatletter
\newcommand*\rel@kern[1]{\kern#1\dimexpr\macc@kerna}
\newcommand*\widebar[1]{%
  \begingroup
  \def\mathaccent##1##2{%
    \rel@kern{0.8}%
    \overline{\rel@kern{-0.8}\macc@nucleus\rel@kern{0.2}}%
    \rel@kern{-0.2}%
  }%
  \macc@depth\@ne
  \let\math@bgroup\@empty \let\math@egroup\macc@set@skewchar
  \mathsurround\z@ \frozen@everymath{\mathgroup\macc@group\relax}%
  \macc@set@skewchar\relax
  \let\mathaccentV\macc@nested@a
  \macc@nested@a\relax111{#1}%
  \endgroup
}
\makeatother

\DeclareFontFamily{U}{mathx}{\hyphenchar\font45}
\DeclareFontShape{U}{mathx}{m}{n}{
	<5> <6> <7> <8> <9> <10>
	<10.95> <12> <14.4> <17.28> <20.74> <24.88>
	mathx10
}{}
\DeclareSymbolFont{mathx}{U}{mathx}{m}{n}
\DeclareFontSubstitution{U}{mathx}{m}{n}

\allowdisplaybreaks

\theoremstyle{remark}

\newtheoremstyle{mytheoremstyle} 
    {\topsep}                    
    {\topsep}                    
    {\upshape}                   
    {.5em}                           
    {\itshape}                   
    {.}                          
    {.5em}                       
    {}  
    
\theoremstyle{mytheoremstyle}

\newtheoremstyle{iremark}
  {\topsep}   
  {\topsep}   
  {\upshape}  
  {0.2in}       
  {\itshape}  
  {.}         
  {5pt plus 1pt minus 1pt} 
  {\thmname{#1}\thmnumber{ \itshape#2}\thmnote{ (#3)}} 

\theoremstyle{iremark}

\newtheorem*{proof}{Proof}

\DeclarePairedDelimiter\abs{\lvert}{\rvert}%
\DeclarePairedDelimiter\absbig{\Big\lvert}{\Big\rvert}%

\makeatletter \renewcommand\d[1]{\ensuremath{%
		\;\mathrm{d}#1\@ifnextchar\d{\!}{}}}
\makeatother

\include{commands}

\graphicspath{{./Figures/}}

\begin{document}
\bstctlcite{IEEEexample:BSTcontrol}
\title{Radar Sensing with OTFS: Embracing ISI and ICI to Surpass the Ambiguity Barrier}
\author{
{Musa Furkan Keskin}\IEEEauthorrefmark{1},
{Henk Wymeersch}\IEEEauthorrefmark{1},
{Alex Alvarado}\IEEEauthorrefmark{2}\\
\IEEEauthorrefmark{1}Chalmers University of Technology, Sweden 
\IEEEauthorrefmark{2}Eindhoven University of Technology, The Netherlands 
}

\maketitle

\begin{abstract}
    Orthogonal time frequency space (OTFS) is a promising alternative to orthogonal frequency division multiplexing (OFDM) in high-mobility beyond 5G communications. In this paper, we consider the problem of radar sensing with OTFS joint radar-communications waveform and derive a novel OTFS radar signal model by explicitly taking into account the inter-symbol interference (ISI) and inter-carrier interference (ICI) effects. On the basis of the new model, we show how ISI and ICI phenomena can be turned into an advantage to surpass the maximum unambiguous detection limits in range and velocity, arising in existing OFDM and OTFS radar systems. Moreover, we design a generalized likelihood ratio test based detector/estimator that can \emph{embrace} ISI and ICI effects. Simulation results illustrate the potential of embracing ISI/ICI and demonstrate its superior detection and estimation performance over conventional baselines.

	\textit{Index Terms--} OTFS, OFDM, joint radar-communications, inter-symbol interference, inter-carrier interference.
\end{abstract}

\section{Introduction}

As 5G systems are being rolled out, the time has come to conceive and develop Beyond 5G (B5G) communication systems. There are now several initiatives in Europa, the USA, and Asia to define what B5G will be in terms of use cases and requirements \cite{saad2019vision,rajatheva2020white,yang20196g}. As with all previous generations, one requirement will be a 10-fold increase in peak data rate. Different from previous generations is the increased emphasis on joint radar and communication (JRC) \cite{jointRadCom_review_TCOM}, driven not only by localization use cases but also the inherent geometric nature of the wireless propagation channel \cite{Lima6Gsensing20,wymeersch2020radio}. 

In the pursuit of higher data rates, 
lower latency, and higher sensing accuracies, we have no choice but to consider larger carrier frequencies, above the 24 GHz band in 5G, as this is where larger bandwidths are available \cite{elayan2018terahertz}. At lower frequencies, (despite intense competition) OFDM has remained the communication waveform of choice, due to its robustness to multipath, its straightforward integration with multi-antenna systems, and its high flexibility in terms of power and rate allocation \cite{banelli2014modulation}. In addition, OFDM is suitable for JRC with standard FFT-processing, in both mono-static and bi-static configurations \cite{braun2014ofdm}. However, at B5G frequencies, OFDM is challenged by several effects, which force us to consider alternative modulation formats: OFDM suffers from a high peak-to-average power ratio (PAPR), leading to reduced  power efficiency, which becomes a limiting factor at high carriers. Secondly, OFDM require frequent adaptation due to mobility and fading, which would lead to prohibitive overheads at high carriers due to the short coherence times \cite{banelli2014modulation}. Moreover, the robustness to multipath comes at a cost of inserting a cyclic prefix (CP) between OFDM symbols, resulting in a rate loss, to combat inter-symbol interference (ISI). Finally, for radar, OFDM is sensitive to inter-carrier-interference (ICI), resulting from Doppler shifts under high target velocities \cite{wang2006performance}. 

These perceived drawbacks of OFDM have renewed the interest in alternative modulation formats, in particular with favorable properties in terms of JRC performance. Orthogonal time frequency space (OTFS) has become a promising candidate in this respect, as it has lower PAPR \cite{surabhi2019peak}, requires less frequent adaptation \cite{hadani2017orthogonal}, has a much lower cyclic overhead \cite{raviteja2018practical}, and can cope with much larger Doppler shifts \cite{murali2018otfs}. In contrast to OFDM's 2D modulation in the time--frequency domain, OTFS relies on the delay--Doppler domain. This implies that a time-varying channel with constant Doppler will appear time-invariant to OTFS \cite{wiffen2018comparison}. Since radar detections are of the form of range (delay) and velocity (Doppler) tuples, OTFS is a natural candidate for JRC, as evidenced by recent activity in this area \cite{OTFS_2019,Gaudio_MIMO_OTFS_Hybrid,otfs_radar_2019,OTFS_RadCom_TWC_2020}. Target detection and estimation of their range and velocity is done by converting the received time-domain signal back to the delay-Doppler domain. This leads to high complexity, e.g., requiring iterative interference cancellation-based processing due to significant side-lobe levels \cite{Gaudio_MIMO_OTFS_Hybrid}. In addition, the expression of the signal in the delay-Doppler domain is complicated (see, e.g., \cite[Eq.~(12)]{OTFS_RadCom_TWC_2020} and \cite[Eq.~(11)]{Gaudio_MIMO_OTFS_Hybrid}), making it difficult to derive insights into the structure of the OTFS signal in terms of ISI and ICI effects. Finally, the existing approaches are limited in terms of ISI and ICI due to standard ambiguity limits in \cite{Gaudio_MIMO_OTFS_Hybrid,otfs_radar_2019,OTFS_RadCom_TWC_2020}. 


\begin{figure}
	\centering
	\includegraphics[width=1\linewidth]{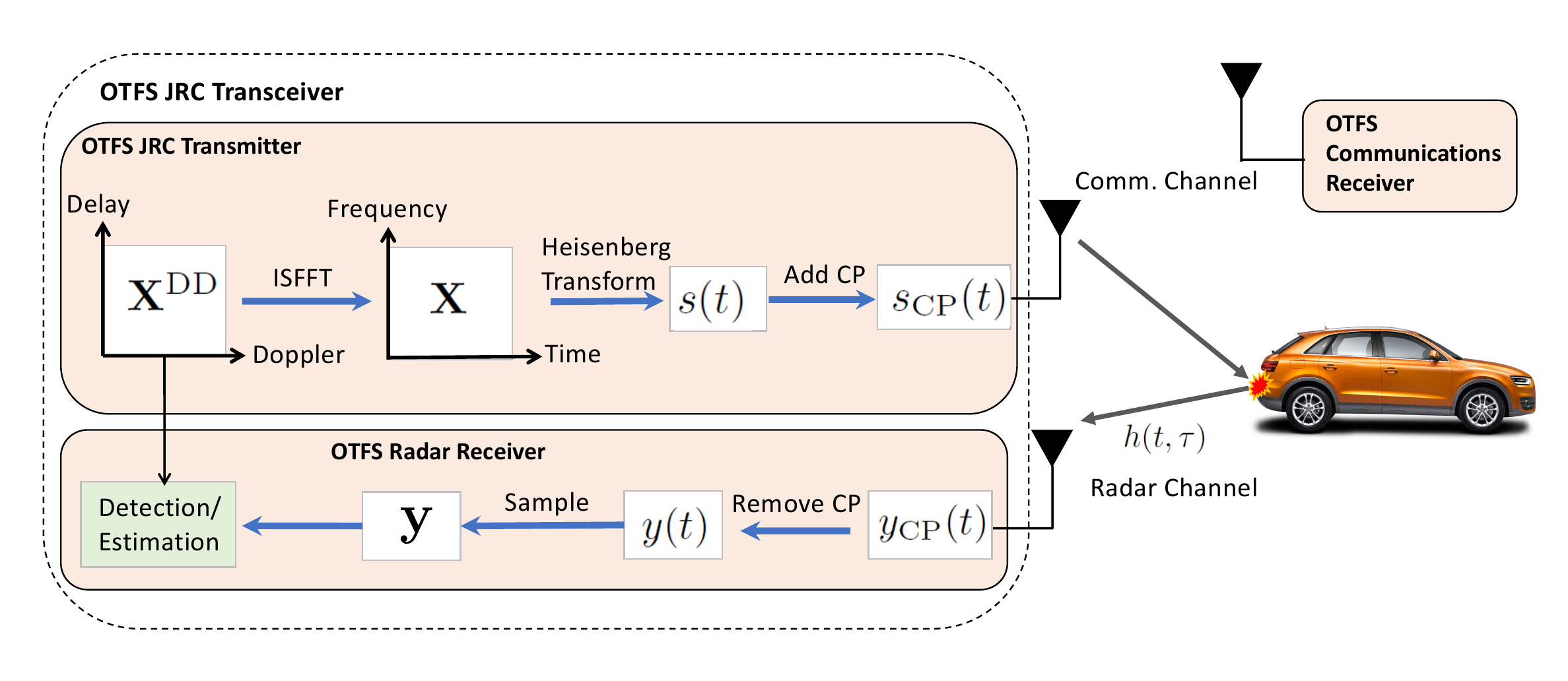}
	\caption{OTFS JRC system, where the delay-Doppler data $\boldXdd \in \complexset{N}{M}$ is converted to the time-frequency domain, and then to the time-domain, where pulse shaping occurs and a single CP is added to the entire frame. The signal $y(t)$ is sampled in the time domain at $t = \ell T/N$ for $\ell = 0, \ldots, NM-1$ and the frequency domain at $f = n \deltaf/M$ for $n = 0, \ldots, NM-1$.}
	\label{fig_system}
\end{figure}

In this paper, we
pursue a different and computationally more appealing approach where the time-domain observations are directly utilized to estimate target parameters without transformation from {time} to {delay-Doppler domain}. This reveals the explicit 
dependency of the observation on target delays and Dopplers, which facilitates detector/estimator design, leading to a low-complexity multi-target generalized likelihood ratio test (GLRT) detector. 
In addition, the proposed time-domain processing approach provides insights into the OTFS waveform structure and thus enables exploitation of the ISI and ICI effects to surpass the standard ambiguity limits. 
%
The main contributions of this paper can be summarized as follows:
\begin{itemize}
    \item We derive a novel time-domain signal model for OTFS radar by explicitly taking into account the ISI and ICI effects. This allows detection of targets directly using the time-domain observations in a low-complexity manner without transformation to delay-Doppler domain and enables turning ISI/ICI into an advantage for sensing.
    \item Based on the new model, we propose an ISI/ICI embracing approach to surpass the range/velocity ambiguity barrier encountered in existing OFDM \cite{RadCom_Proc_IEEE_2011,OFDM_Radar_Phd_2014} and OTFS  \cite{Gaudio_MIMO_OTFS_Hybrid,otfs_radar_2019,OTFS_RadCom_TWC_2020} radar algorithms, which enables detection of any practically relevant range/velocity.
    \item We design a generalized likelihood ratio test (GLRT) based detector/estimator for OTFS radar that simultaneously compensates for the ISI/ICI and harnesses their properties to improve performance. 
\end{itemize}



\section{OTFS Radar Signal Model}
We consider an OTFS JRC system consisting of an OTFS JRC transceiver and an OTFS communications receiver, as shown in Fig.~\ref{fig_system}. The OTFS transceiver is equipped with an OTFS transmitter that generates the JRC signal for communicating with the OTFS communication receiver and an OTFS radar receiver that processes the backscattered signals for target detection. In this section, we derive OTFS radar signal model for the single-CP OTFS modulation architecture from \cite{reducedCP_OTFS_2018,otfs_radar_2019,otfs_frac_2020,OTFS_RadCom_TWC_2020}. As our contribution is not on OTFS communication, the processing at the OTFS communication receiver will be ignored. 





The OTFS system has a total bandwidth $N \deltaf$ and total frame duration $M T$ (excluding any CP), where $N$ and $M$ denote the number of subcarriers and the number of symbols, respectively, $\deltaf$ is the subcarrier spacing and $T = 1/\deltaf$ represents the symbol duration. 
%
Let $\boldXdd \in \complexset{N}{M}$ denote the two-dimensional (2-D) OTFS frame in the delay-Doppler domain consisting of $NM$ transmit data symbols that reside on the delay-Doppler grid (see Fig.~\ref{fig_system}) 
\begin{equation}\nonumber
    \mathcal{G} = \left\{ \left(\frac{n}{N \deltaf}, \frac{m}{MT}\right) ~\Big\lvert ~ 0 \leq n \leq N-1, 0 \leq m \leq M-1 \right\}~.
\end{equation}
Applying an inverse symplectic finite Fourier transform (ISFFT) (i.e., an $N$-point FFT over the columns and an $M$-point IFFT over the rows of $\boldXdd$), we transform the 2-D transmit data block from the delay-Doppler domain to the frequency-time domain \cite{hadani2017orthogonal,OTFS_CE_TSP_2019,reducedCP_OTFS_2018,OFDM_OTFS_modem_2017,OTFS_Canc_Iterative_TWC_2018}
\begin{equation}\label{eq_otfs_dd2ft}
    \boldX = \FF_N \boldXdd \FF_M^H ~, 
\end{equation}
where $\boldX \in \complexset{N}{M}$ is the frequency-time domain signal and $\FF_N \in \complexset{N}{N}$ is the unitary DFT matrix with $\left[ \FF_N\right]_{\ell,n} = \frac{1}{\sqrt{N}} e^{- j 2 \pi n \frac{\ell}{N}} $.

To map the frequency-time domain 2-D sequence $\boldX$ to a time domain signal transmitted over the wireless channel, we apply the \textit{Heisenberg transform} \cite{hadani2017orthogonal,OTFS_Canc_Iterative_TWC_2018}, which entails an $N$-point IFFT together with a transmit pulse-shaping waveform $\gtx(t)$ (which is time limited to $\left[0, \, T \right]$). The time domain signal corresponding to the $\thnew{m}$ symbol after the Heisenberg transform can be written as
\begin{equation}\label{eq_smt}
    s_m(t) = \frac{1}{\sqrt{N}} \sum_{n = 0}^{N-1}  \xnm \, e^{j 2 \pi n \deltaf t} \gtx(t) \,, ~ 0 \leq t \leq T ~,
\end{equation}
where $\xnm \triangleq \left[ \boldX \right]_{n,m}$. Hence, the time domain signal for the entire OTFS frame without CP is given by
\begin{equation}\label{eq_st}
    s(t) = \sum_{m=0}^{M-1} s_m(t-mT) \, , ~ 0 \leq t \leq MT ~.
\end{equation}
%
Finally, the entire time domain signal with CP is given by 
\begin{equation}\label{eq_smaone_baseband}
\scp(t) = \begin{cases} s(t) ,&~~ 0 \leq t \leq MT \\
s(t+MT) ,&~~ -\Tcp \leq t \leq 0 \end{cases}  ~,
\end{equation} 
where $\Tcp$ denotes the CP duration.


We consider a $K$-tap doubly selective, narrowband radar channel model as $h(t, \tau) = \sum_{k=0}^{K-1} \alpha_k \delta(\tau - \tau_k)  e^{j 2 \pi \nu_k t}$ \cite{80211_Radar_TVT_2018,OTFS_RadCom_TWC_2020}, 
where the $\thnew{k}$ target is characterized by a complex channel gain $\alpha_k$, a round-trip delay $\tauk = 2 R_k/c$ and a Doppler shift $\nuk = 2  \vk/ \lambda$, with $R_k$, $\vk$, $c$ and $\lambda$ denoting the range, radial velocity, speed of propagation and carrier wavelength, respectively.

After transmission through the channel, the backscattered signal at the OTFS radar receiver can be expressed as
\begin{align}\label{eq_yt}
    \ycp(t) 
    &     = \sum_{k=0}^{K-1} \alpha_k \scp(t - \tau_k)  e^{j 2 \pi \nu_k t}+w(t)
\end{align}
for $-\Tcp \leq t \leq M T$, where $\scp(t)$ is given by \eqref{eq_smaone_baseband} and $w(t)$ is additive white Gaussian noise. We assume that the CP duration is larger than the round-trip delay of the furthermost target, i.e., $\Tcp \geq \max_{k\in \{0,\ldots,K-1\}} \tau_k$.

\section{Novel OTFS Signal Model for ISI/ICI Exploitation}\label{sec_ma1_model}
In this section, we derive a novel compact representation of the received signal in \eqref{eq_yt} for OTFS radar and show how it can be exploited to improve radar performance. 

\subsection{Derivation of Received Signal for OTFS Radar}\label{sec_der_rec_otfs}
We first remove the CP in \eqref{eq_yt} (i.e., the interval $\left[-\Tcp, \, 0 \right]$) to obtain
\begin{equation}\label{eq_yt_baseband}
y(t) = \begin{cases} \ycp(t) ,&~~ 0 \leq t \leq MT \\
0 ,&~~ -\Tcp \leq t \leq 0 \end{cases}  ~.
\end{equation} 
Under the assumption $\Tcp \geq \max_k \tau_k$, the signal in \eqref{eq_yt_baseband} becomes a superposition of cyclically shifted copies of the transmit signal $s(t)$ in \eqref{eq_st} \cite[eq.~(6)]{reducedCP_OTFS_2018}:
\begin{equation}\label{eq_yt_ma1_cp}
        y(t) = \sum_{k=0}^{K-1} \alpha_k s([t - \tau_k]_{MT})  e^{j 2 \pi \nu_k t}  +w(t) ~, 
\end{equation}
where $[\cdot]_T$ denotes modulo-$T$. 

Let $S(f) \triangleq \mathcal{F}\{s(t)\} = \int_{0}^{MT} s(t) e^{-j 2 \pi f t}\d t$ denote the Fourier transform of $s(t)$. Then, a cyclic shift of $s(t)$ corresponds to a phase shift in $S(f)$, i.e.,
\begin{equation}
    s([t - \tau_k]_{MT}) = \mathcal{F}^{-1}\big\{ S(f) e^{-j 2 \pi f \tau_k} \big\} ~,
\end{equation}
where $\mathcal{F}^{-1}\{\cdot\}$ represents the inverse Fourier transform. Sampling the time domain at $t = \ell T/N$ for $\ell = 0, \ldots, NM-1$ and the frequency domain at $f = n \deltaf/M$ for $n = 0, \ldots, NM-1$, the equivalent discrete-time representation of \eqref{eq_yt_ma1_cp} can be written as
\begin{equation}\label{eq_yt_ma1_cp_discrete}
    \yy = \sum_{k=0}^{K-1} \alpha_k \,  \FF^H \big( \FF \sss \odot \bb(\tau_k) \big)  \odot \cc(\nu_k) + \ww ~,
\end{equation}
 where $\odot$ is the Hadamard (element-wise) product, $\FF \in \complexset{NM}{NM}$ is the unitary DFT matrix, $\sss \in \complexset{NM}{1}$ and $\yy \in \complexset{NM}{1}$ denote the sampled versions of $s(t)$ and $y(t)$, respectively, $\ww \in \complexset{NM}{1}$ is the additive white Gaussian noise with $\ww \sim \mtCN(\boldzero, \sigma^2 \Imatrix )$, 
\begin{equation}\label{eq_b_steer}
    \bb(\tau) = \bb_{N}(\tau) \otimes \bb_{M}(\tau) \in \complexset{NM}{1}
\end{equation}
is the \textit{frequency-domain steering vector} with
\begin{align*}
    \bb_{M}(\tau) &\triangleq  \transpose{ \left[ 1, e^{-j 2 \pi \frac{1}{M} \deltaf \tau}, \ldots,  e^{-j 2 \pi \frac{M-1}{M} \deltaf  \tau} \right] } \in \complexset{M}{1} ~, \\
    \bb_{N}(\tau) &\triangleq  \transpose{ \left[ 1, e^{-j 2 \pi \deltaf \tau}, \ldots,  e^{-j 2 \pi (N-1) \deltaf  \tau} \right] } \in \complexset{N}{1} ~,
\end{align*}
and, 
\begin{equation}\label{eq_c_steer}
    \cc(\nu) = \cc_{M}(\nu) \otimes \cc_{N}(\nu) \in \complexset{NM}{1}    
\end{equation}
is the \textit{temporal steering vector} with
    \begin{align*}
        \cc_M(\nu) & \triangleq \transpose{ \left[ 1, e^{j 2 \pi T \nu }, \ldots,  e^{j 2 \pi (M-1) T \nu } \right] } \in \complexset{M}{1} ~, \\
        \cc_N(\nu) &\triangleq \transpose{ \left[ 1, e^{j 2 \pi  \frac{T}{N} \nu}, \ldots, e^{j 2 \pi \frac{T(N-1)}{N} \nu}  \right]   } \in \complexset{N}{1} ~.
    \end{align*}
    Here,  $\otimes$ denotes the  Kronecker product.
The vectors $\bb_{N}(\tau)$ and $\cc_M(\nu)$ are commonly encountered in OFDM radar, used for recovering the range and velocity, respectively. In contrast, the vectors $\bb_{M}(\tau)$ and $\cc_N(\nu)$ are commonly seen as disturbances that degrade the performance. 
In particular, $\bb_{N}(\tau)$ quantifies delay-dependent frequency-domain phase rotations corresponding to the Fourier transform of fast-time (hereafter called \textit{fast-frequency} domain), while $\bb_{M}(\tau)$ involves inter-symbol (slow-time) delay-dependent phase rotations (hereafter called \textit{slow-frequency} domain), leading to \textit{inter-symbol interference (ISI)}. Similarly, $\cc_N(\nu)$ represents Doppler-induced fast-time phase rotations causing \textit{inter-carrier interference (ICI)}, similar to the carrier frequency offset (CFO) effect in OFDM communications \cite{Visa_CFO_TSP_2006}, while $\cc_M(\nu)$ captures Doppler-dependent slow-time phase progressions. 

\subsection{Increasing Unambiguous Detection Intervals via ISI and ICI Exploitation}\label{sec_isi_ici}
The steering vector structures in \eqref{eq_b_steer} and \eqref{eq_c_steer} suggest that radar receivers in  OTFS suffer from both ISI and ICI effects \cite{OTFS_RadCom_TWC_2020}, similar to the case of OTFS communications \cite{OTFS_Canc_Iterative_TWC_2018}. However, unlike OTFS communications, these two effects can be turned into an advantage for OTFS radar. In particular, ISI manifests itself through the \textit{slow-frequency} steering vector $\bb_{M}(\tau)$ and enables sampling the available bandwidth $N \deltaf$ at integer multiples of $\deltaf / M$ as observed from the Kronecker structure in \eqref{eq_b_steer}. In ISI-free operation, the steering vector in \eqref{eq_b_steer} would only involve the \textit{fast-frequency} component $\bb_{N}(\tau)$, which can sample the bandwidth with a spacing of $\deltaf$, i.e., the subcarrier spacing. Hence, ISI can increase the maximum detectable unambiguous delay by a factor of $M$ by sampling the frequency domain $M$ times faster compared to a standard ISI-free radar operation (e.g., in OFDM-based OTFS radar\footnote{Contrary to single-CP OTFS, OFDM-based OTFS systems use separate CPs for each symbol in the OTFS/OFDM frame to circumvent the ISI effect \cite{OTFS_CE_TSP_2019,otfs_modem_2018}.}). More precisely, the unambiguous delays with and without ISI are given, respectively, by
\begin{align}\label{eq_isi_exp}
    \taumaxisi = \min \Bigg\{ \frac{M}{\deltaf}, \Tcp \Bigg\}, ~ \taumax = \min \Bigg\{ \frac{1}{\deltaf}, \Tcp \Bigg\} ~,
\end{align}
where $\Tcp$ is the upper limit to prevent inter-frame interference (i.e., between consecutive OTFS frames) in  OTFS radar and to prevent ISI between consecutive symbols in OFDM-based OTFS radar.

Similarly to ISI, ICI can be exploited to increase the maximum detectable unambiguous Doppler by a factor of $N$ \cite{MIMO_OFDM_ICI_JSTSP_2021}. In addition to the standard slow-time steering vector $\cc_M(\nu)$, Doppler-dependent phase rotations can also be captured by the \textit{fast-time} steering vector $\cc_N(\nu)$, which allows sampling the entire time window $M T$ with an interval of $T/N$. Therefore, the unambiguous Doppler values with and without ICI can be expressed, respectively, as
\begin{align}\label{eq_ici_exp}
    \numaxici =\frac{N}{T}, ~ \numax =  \frac{1}{T} ~.
\end{align}

\section{Detection and Estimation with OTFS Radar}
In this section, we provide the problem statement for OTFS radar sensing and design a low-complexity detection and estimation scheme based on the signal model in Sec.~\ref{sec_der_rec_otfs}.

\subsection{Problem Formulation for OTFS Radar Sensing}\label{sec_otfs_prob}
Given the transmit signal $\sss$, the problem of interest for OTFS radar sensing is to detect the presence of multiple targets and estimate their parameters, i.e., their gain-delay-Doppler triplets $\{(\alpha_k, \tau_k, \nu_k)\}_{k=0}^{K-1}$. This detection and estimation is done from the time domain observations in \eqref{eq_yt_ma1_cp}, or, equivalently, from its sampled version in \eqref{eq_yt_ma1_cp_discrete}. Unlike the existing works in the OTFS radar literature, e.g., \cite{Gaudio_MIMO_OTFS_Hybrid,otfs_radar_2019,OTFS_RadCom_TWC_2020}, where estimator design is based on the received symbols in the delay-Doppler domain, we adopt a low-complexity approach that performs detection/estimation directly using \textit{time-domain} observations without transforming them into \textit{delay-Doppler domain}. 


\subsection{GLRT for Detection/Estimation in OTFS Radar}\label{sec_grlt}
To design a detector, we first rewrite the signal model in \eqref{eq_yt_ma1_cp_discrete} as
\begin{align}\label{eq_y_new}
    \yy = \sum_{k=0}^{K-1} \alpha_k \, \boldC(\nu_k)  \FF^H \boldB(\tau_k) \FF \sss  + \ww ~,
\end{align}
where $\boldB(\tau) \triangleq \diag{\bb(\tau)} \in \complexset{NM}{NM}$ and $\boldC(\nu) \triangleq \diag{\cc(\nu)} \in \complexset{NM}{NM}$. The hypothesis testing problem to test the presence of a single target in \eqref{eq_y_new} can be expressed as
\begin{align}\label{eq_hypo}
    \yy = \begin{cases}
	\ww,&~~ {\rm{under~\mathcal{H}_0}}  \\
	\alpha \, \boldC(\nu)  \FF^H \boldB(\tau) \FF \sss  + \ww ,&~~ {\rm{under~\mathcal{H}_1}} 
	\end{cases} ~,
\end{align}
where the hypotheses $\mathcal{H}_0$ and $\mathcal{H}_1$ represent the absence and presence of a target, respectively. To solve \eqref{eq_hypo}, we treat $\alpha$, $\tau$ and $\nu$ as deterministic unknown parameters and resort to the GLRT
\begin{equation}\label{eq_glrt}
    \Lambda(\yy) = \frac{ \max_{\alpha, \tau, \nu} p(\yy \, \lvert \, \mathcal{H}_1 ; \alpha, \tau, \nu ) }{p(\yy \, \lvert \, \mathcal{H}_0  ) } \underset{\mathcal{H}_0}{\overset{\mathcal{H}_1}{\gtrless}} \eta~,
\end{equation}
where $\eta$ is the threshold. Under the assumption $\ww \sim \mtCN(\boldzero, \sigma^2 \Imatrix )$, the GLRT becomes
\begin{align}\label{eq_glrt2}
    & \Lambda(\yy) \\ \nonumber & = \frac{ \exp\left(- \frac{1}{\sigma^2} \min_{\alpha, \tau, \nu} \norm{  \yy - \alpha \, \boldC(\nu)  \FF^H \boldB(\tau) \FF \sss  }^2 \right)  }{\exp\left(- \frac{1}{\sigma^2}  \norm{  \yy }^2  \right) } \underset{\mathcal{H}_0}{\overset{\mathcal{H}_1}{\gtrless}} \eta ~.
\end{align}
For fixed $\tau$ and $\nu$, the optimal channel gain that maximizes the numerator in \eqref{eq_glrt2} is given by
\begin{align} \label{eq_alpha_hat}
    \alpha = \frac{ \sss^H \FF^H \boldB^H(\tau) \FF \boldC^H(\nu) \yy  }{ \norm{\sss}^2 }~.
\end{align}
Plugging \eqref{eq_alpha_hat} back into \eqref{eq_glrt2} and taking the logarithm, we have the detection test
\begin{equation}\label{eq_glrt_final}
    \max_{\tau, \nu} \frac{ \absbig{\sss^H \FF^H \boldB^H(\tau) \FF \boldC^H(\nu) \yy}^2 }{ \sigma^2 \norm{\sss}^2 } \underset{\mathcal{H}_0}{\overset{\mathcal{H}_1}{\gtrless}} \etatilde ~,
\end{equation}
where $\etatilde = \log \eta$. To account for the presence of multiple targets, the decision statistic in \eqref{eq_glrt_final} can be computed over a discretized delay-Doppler region and targets are declared at those locations where there is a peak exceeding the threshold \cite[Ch.~6.2.4]{richards2005fundamentals}.

\section{Numerical Results}
In this section, we assess the performance of the proposed 2-D GLRT based detector/estimator in \eqref{eq_glrt_final}. The OTFS radar observations are generated using \eqref{eq_yt} instead of the proposed model in \eqref{eq_yt_ma1_cp_discrete} to provide an implicit verification of the transition from \eqref{eq_yt} to \eqref{eq_yt_ma1_cp_discrete}. As a benchmark, we consider a standard 2-D FFT based processing traditionally employed in OFDM radar \cite{RadCom_Proc_IEEE_2011,OFDM_Radar_Phd_2014}. 
To evaluate detection performances for both GLRT and FFT methods, a cell-averaging CFAR detector is employed with the probability of false alarm $\pfa = 10^{-4}$ to declare targets in the delay-Doppler domain. For a target with channel gain $\alpha$, we define the signal-to-noise ratio (SNR) as $\snr = \abs{\alpha}^2/\sigma^2$. In addition, a rectangular pulse shaping waveform is used for $\gtx(t)$ in \eqref{eq_smt}.
We consider two different parameter sets for OTFS, as shown in Table~\ref{tab_parameters}, to illustrate the results in both ISI-dominant (i.e., high $\deltaf$) and ICI-dominant (i.e., small $\deltaf$) operation regimes. In the ISI-dominant regime, maximum range is the limiting factor for radar detection performance, while in the ICI-dominant regime, radar performance is mainly limited by maximum velocity. 


\begin{table}
\caption{OTFS Parameter Sets for Simulations}
\centering
    \begin{tabular}{|l|l|l|}
        \hline
        \textbf{Parameter} & \textbf{ISI-dominant} & \textbf{ICI-dominant} \\ 
         & \textbf{Regime} & \textbf{Regime} \\ \hline
        Carrier Frequency, $\fc$ & $60 \, \rm{GHz}$ & $60 \, \rm{GHz}$ \\ \hline
        Subcarrier Spacing, $\deltaf$ & $781.3 \, \rm{kHz}$ & $48.8 \, \rm{kHz}$  \\ \hline
        Number of Subcarriers, $N$ & $64$ & $1024$ \\ \hline
        Total Bandwidth, $B$ & $50 \, \rm{MHz}$ & $50 \, \rm{MHz}$ \\ \hline
        Symbol Duration, $T$ & $1.28 \, \rm{\mu s}$ & $20.48 \, \rm{\mu s}$ \\ \hline 
        Cyclic Prefix Duration, $\Tcp$ & $7.68 \, \rm{\mu s}$ & $20.48 \, \rm{\mu s}$ \\ \hline 
        Number of Symbols, $M$ & $64$ & $8$ \\ \hline
        Frame Duration, $M T + \Tcp$ & $89.6 \,  \rm{\mu s}$ & $184.3 \, \rm{\mu s}$ \\ \hline 
                
        Range Resolution, $\deltar$ & $3 \, \rm{m}$ & $3 \, \rm{m}$ \\ \hline
        Maximum Range, $\Rmax$ & $192 \, \rm{m}$ & $3072 \, \rm{m}$ \\ 
        (\textit{Standard}) & &
        \\ \hline
        Maximum Range, $\rmaxisi$ & $1152 \, \rm{m}$ & $3072 \, \rm{m}$ \\ 
        (\textit{ISI Embracing}) & &
        \\ \hline
        Velocity Resolution, $\deltav$ & $30.5 \, \rm{m/s}$ &  $15.3 \, \rm{m/s}$ \\ \hline
        Maximum Velocity, $\vmax$ & $\pm 976.6 \, \rm{m/s}$ & $\pm 61 \, \rm{m/s}$ \\ 
        (\textit{Standard}) & & \\ \hline
        Maximum Velocity, $\vmaxici $ & no practical limit & no practical limit \\
         (\textit{ICI Embracing}) & & \\ \hline
    \end{tabular}
    \label{tab_parameters}
    \vspace{-0.1in}
\end{table}


\subsection{Example 1: ISI-dominant Regime}
\begin{figure}
	\centering
	\includegraphics[width=1\linewidth]{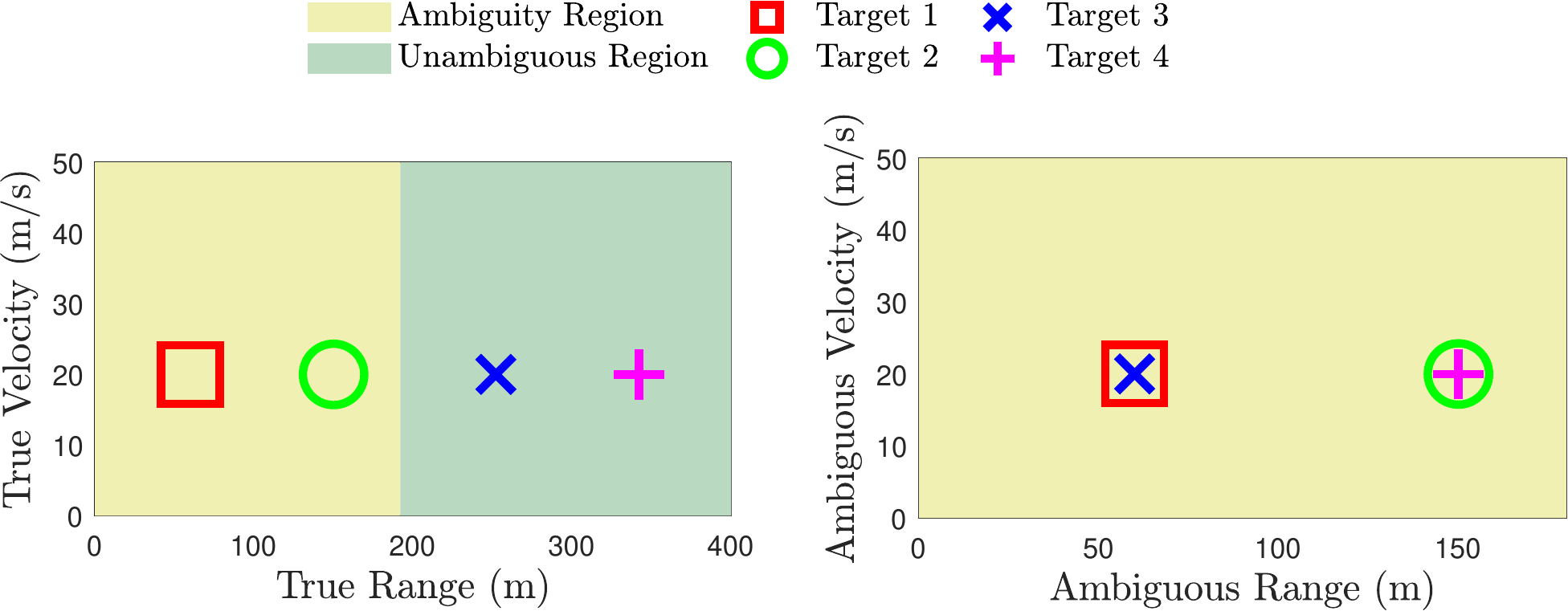}
	\caption{Scenario for OTFS sensing in the ISI-dominant regime, where target SNRs are given by $\{ 25, 10, 20, 10 \} \, \rm{dB}$, respectively.}
	\label{fig_scenario_isi_dominant}
	\vspace{-0.1in}
\end{figure}

In the ISI-dominant regime, we consider a scenario with four targets located at the same velocity ($20$~m/s), but with different ranges, as shown in Fig.~\ref{fig_scenario_isi_dominant} (left). As shown in Fig.~\ref{fig_scenario_isi_dominant} (right), the standard FFT processing \cite{RadCom_Proc_IEEE_2011,OFDM_Radar_Phd_2014} or the existing OTFS detectors \cite{Gaudio_MIMO_OTFS_Hybrid,otfs_radar_2019,OTFS_RadCom_TWC_2020} can detect at most two targets as Target~3 and Target~4 fall into the same range-velocity bin as Target~1 and Target~2, respectively. 

\begin{figure}
        \begin{center}
        \subfigure[]{
			 \label{fig_rp_isi_1}
			 \includegraphics[width=0.4\textwidth]{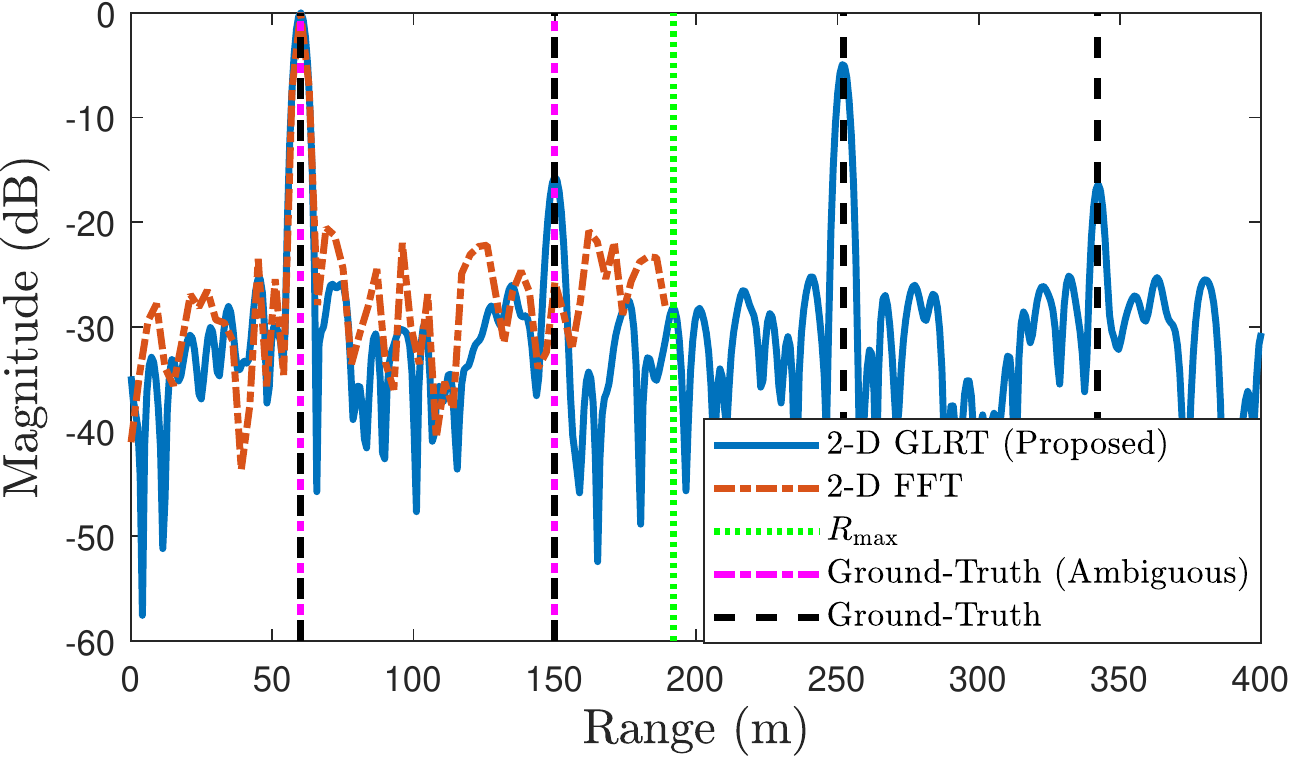}
		}
        \subfigure[]{
			 \label{fig_rp_isi_2}
			 \includegraphics[width=0.4\textwidth]{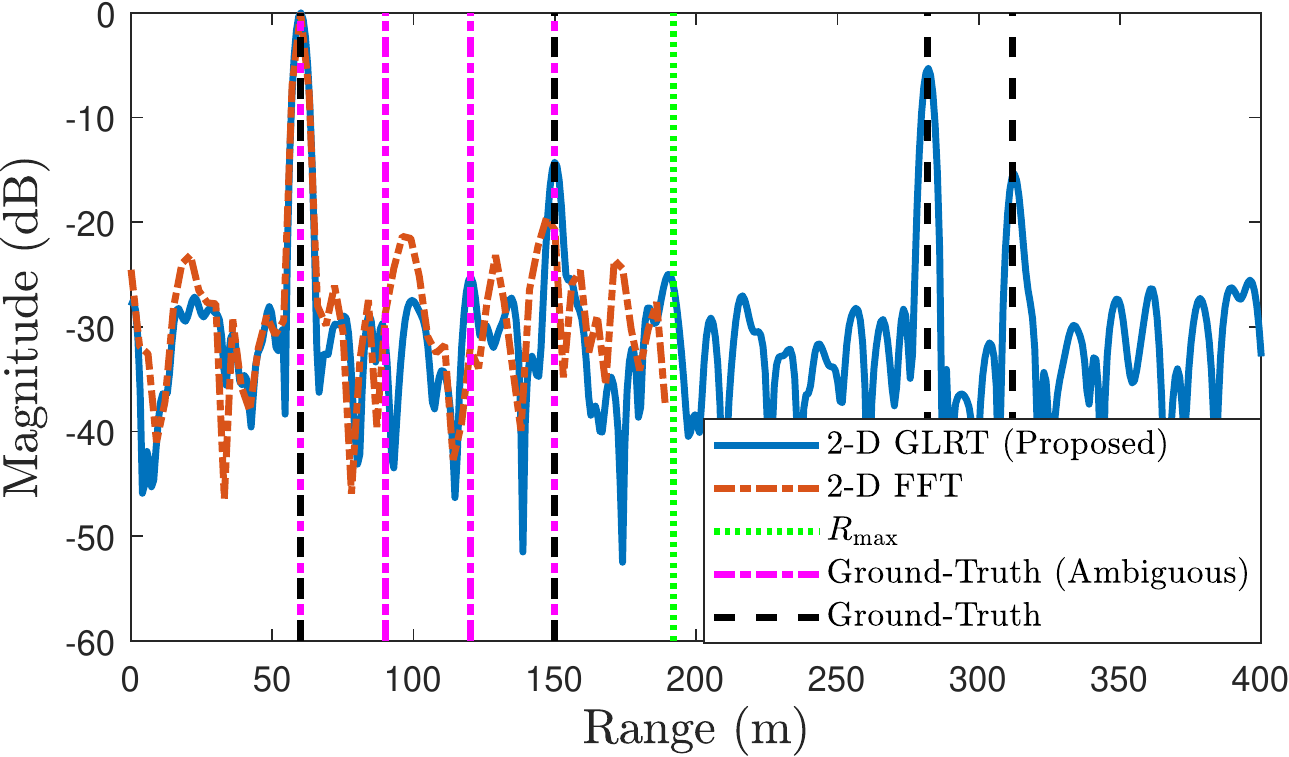}
		}
		
		\end{center}
		\vspace{-0.2in}
        \caption{ISI-dominant regime: Range profile at $v = 20 \, \rm{m/s}$ obtained by the different methods \subref{fig_rp_isi_1} for the scenario in Fig.~\ref{fig_scenario_isi_dominant}, and \subref{fig_rp_isi_2} for a modified version of the scenario in Fig.~\ref{fig_scenario_isi_dominant}, where Target~3 moved $30 \, \rm{m}$ further and Target~4 moved $30 \, \rm{m}$ closer.}  
        \label{fig_rp_isi}
        \vspace{-0.12in}
\end{figure}
Fig.~\ref{fig_rp_isi} shows the range profiles\footnote{The range profile of the 2-D GLRT in \eqref{eq_glrt_final} is obtained by plotting the decision statistic in \eqref{eq_glrt_final} for a fixed Doppler $\nu$ over an interval of delay values $\tau$. For the 2-D FFT method, range profile corresponds to the range slice of the 2-D FFT output, taken from a certain Doppler $\nu$.} of the considered methods in two different scenarios for a single noise realization. It is observed that by virtue of the ISI embracing, the proposed GLRT detector in \eqref{eq_glrt_final} can detect four targets separately by increasing the maximum range by a factor of $6$ (see \eqref{eq_isi_exp} and Table~\ref{tab_parameters}), whereas the 2-D FFT yields a peak only at the location of Target~1. In addition, the GLRT detector achieves lower side-lobe levels than the FFT method by taking into account the ISI in detector design, which enables compensating for its effect on the range profile. Moreover, even when Target~3 and Target~4 are displaced in Fig.~\ref{fig_rp_isi_2} so that four targets are resolvable in the ambiguity region, the FFT can only detect Target~1 due to the strong ISI effect, while all the target peaks are clearly visible in the range profile of the GLRT detector. Therefore, the proposed approach can simultaneously mitigate ISI to have low side-lobes and embrace the information conveyed by ISI to detect targets beyond the standard maximum range limit $\Rmax$.

\begin{figure}
        \begin{center}
        \subfigure[]{
			 \label{fig_pd_snr_isi}
			 \includegraphics[width=0.4\textwidth]{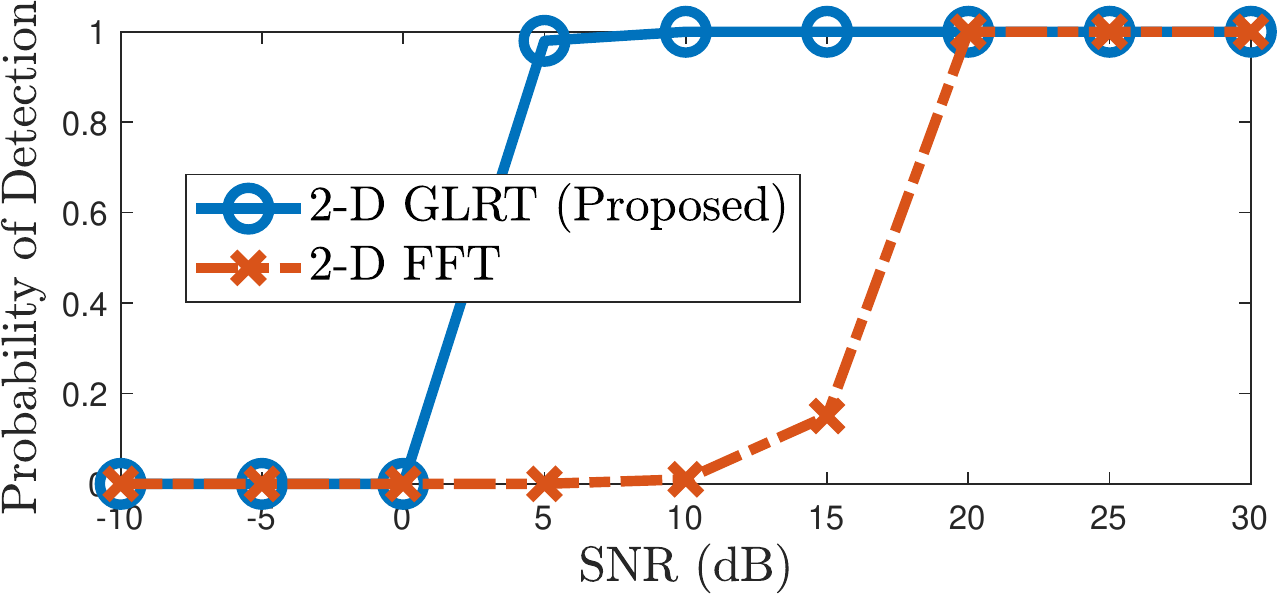}
		}
        \subfigure[]{
			 \label{fig_pfa_snr_isi}
			 \includegraphics[width=0.4\textwidth]{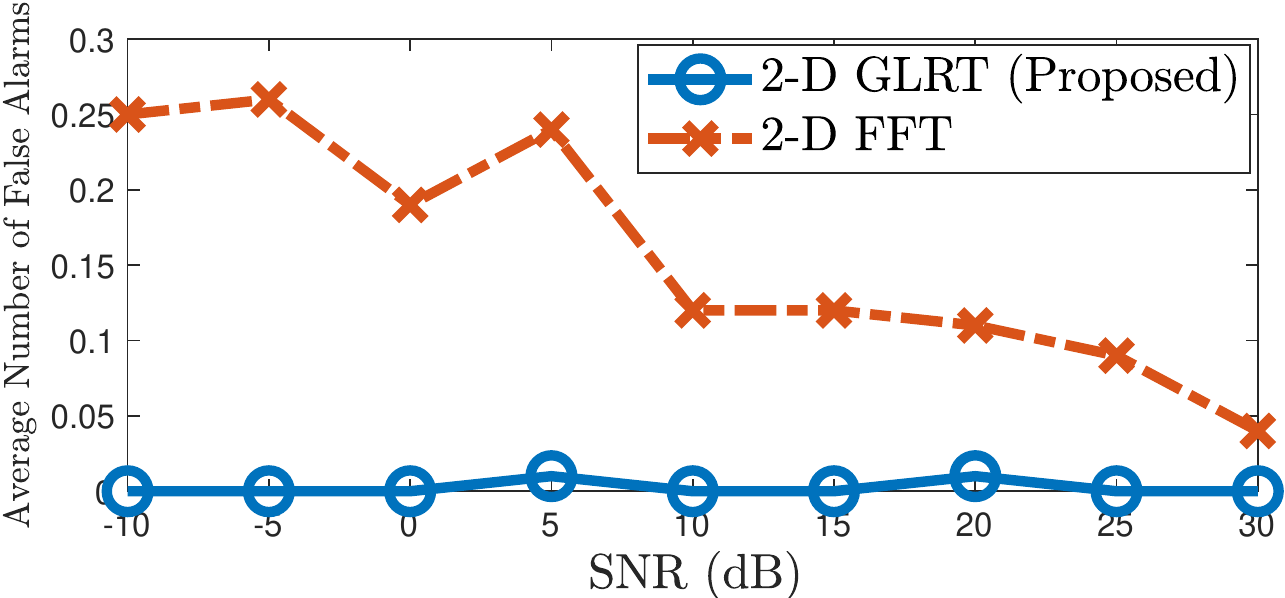}
		}
		
		\subfigure[]{
			 \label{fig_range_rmse_isi}
			 \includegraphics[width=0.4\textwidth]{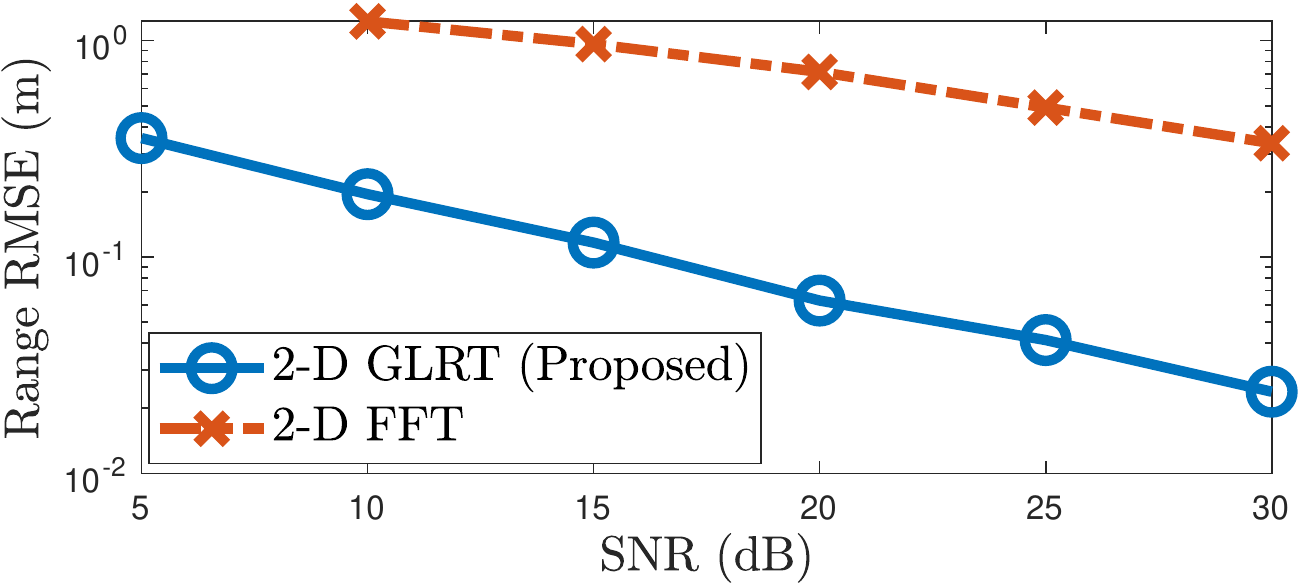}
		}

		\end{center}
		\vspace{-0.2in}
        \caption{Detection and estimation performance of the considered OTFS radar algorithms with respect to SNR in the ISI-dominant regime. \subref{fig_pd_snr_isi} Probability of detection, \subref{fig_pfa_snr_isi} average number of false alarms, and \subref{fig_range_rmse_isi} range RMSE.}  
        \label{fig_isi_det_est}
        \vspace{-0.2in}
\end{figure}

To illustrate the detection and estimation performance, we simulate $100$ independent Monte Carlo noise realizations and choose Target~2 as the reference target. Fig.~\ref{fig_isi_det_est} shows the probability of detection and root mean-squared error (RMSE) of range estimates of the reference target, and the average number of false alarms (detections that do not belong to any of the targets) with respect to the SNR of the reference target. It is seen that the proposed GLRT detector/estimator significantly outperforms the standard FFT method in terms of both detection and estimation performances. This is accomplished through the ISI-aware modeling in \eqref{eq_yt_ma1_cp_discrete} and the corresponding detector design in \eqref{eq_glrt_final}, which performs ISI compensation via the term $\boldB^H(\tau)$. We note that the GLRT detector/estimator in \eqref{eq_glrt_final} performs block-wise processing of the entire OTFS frame ($NM$ symbols), while the FFT method applies separate $N$- and $M$-point FFTs over frequency and time domains, respectively, which provides computational simplicity, but leads to poor radar performance.

\subsection{Example 2: ICI-dominant Regime}
\begin{figure}
	\centering
	\includegraphics[width=1\linewidth]{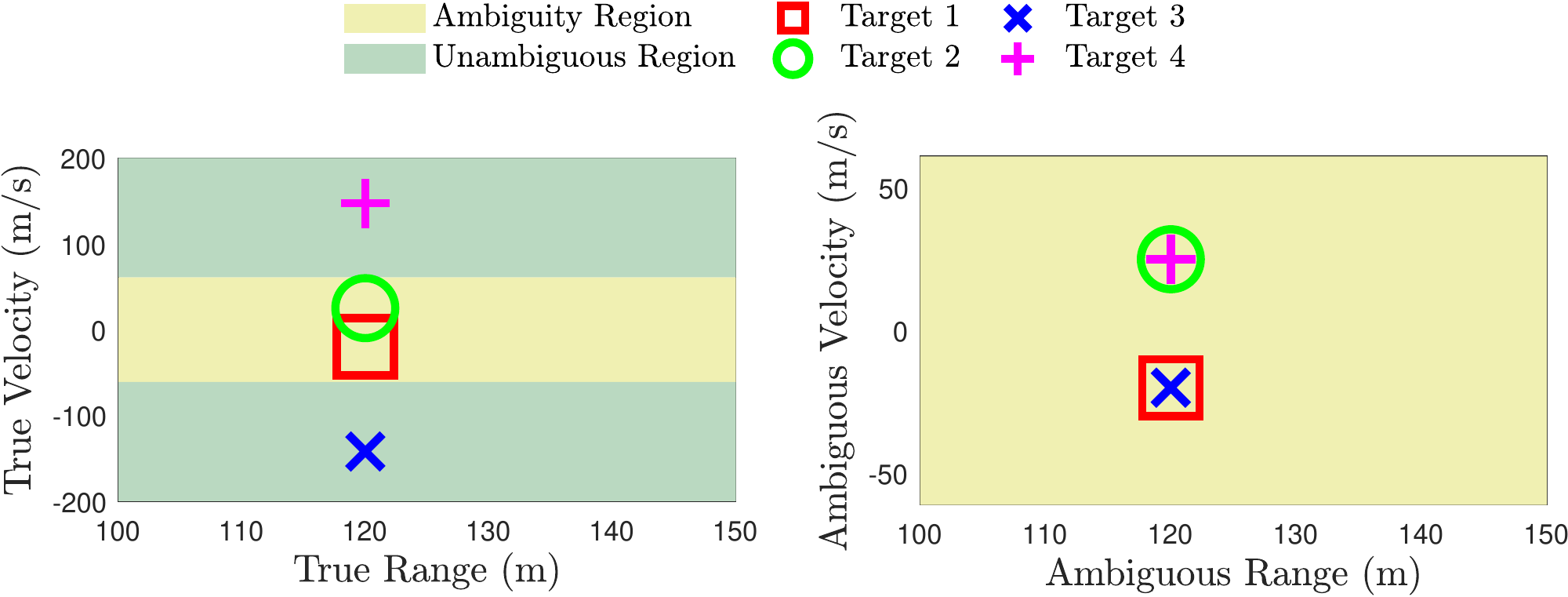}
	\caption{Scenario for OTFS sensing in the ICI-dominant regime, where target SNRs are given by $\{ 25, 10, 20, 10 \} \, \rm{dB}$, respectively.}
	\label{fig_scenario_ici_dominant}
	\vspace{-0.15in}
\end{figure}

For the ICI-dominant regime, the scenario in Fig.~\ref{fig_scenario_ici_dominant} is considered, where four targets are located at the same range bin, but with different velocities. Since the maximum velocity is small, Target~3 and Target~4 lie in the same range-velocity bin as Target~1 and Target~2, respectively. Hence, standard OFDM \cite{RadCom_Proc_IEEE_2011,OFDM_Radar_Phd_2014} and OTFS \cite{Gaudio_MIMO_OTFS_Hybrid,otfs_radar_2019,OTFS_RadCom_TWC_2020} radar algorithms cannot distinguish Target~1-3 and Target~2-4 as two separate echoes.

\begin{figure}
        \begin{center}
        \subfigure[]{
			 \label{fig_vp_ici_1}
			 \includegraphics[width=0.4\textwidth]{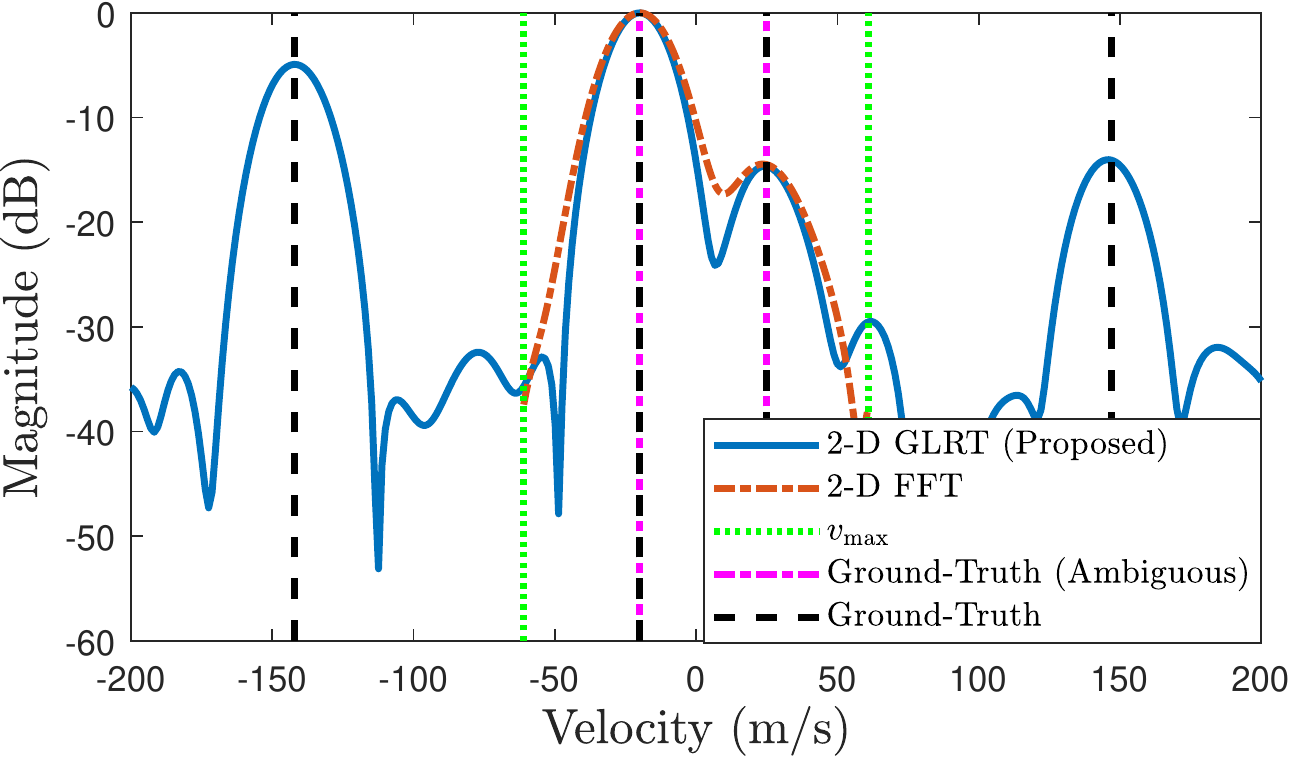}
		}
        \subfigure[]{
			 \label{fig_vp_ici_2}
			 \includegraphics[width=0.4\textwidth]{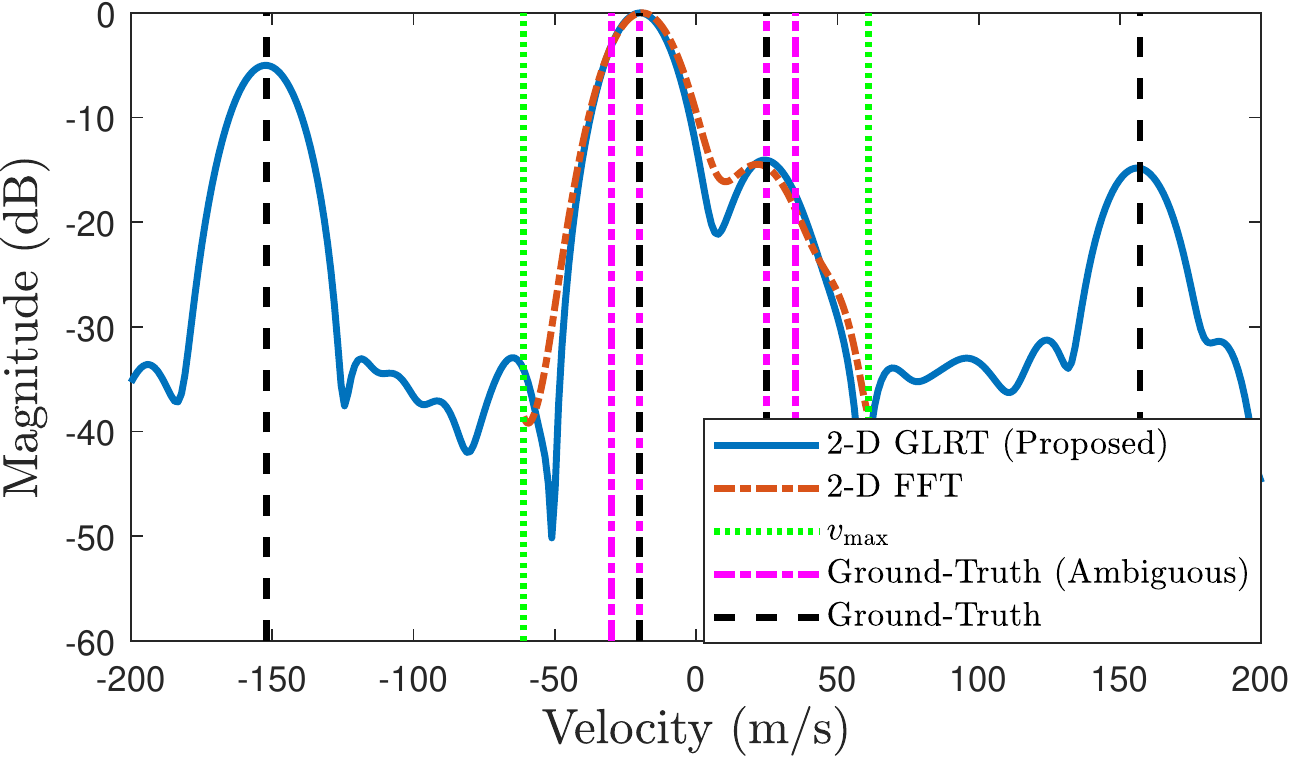}
		}
		
		\end{center}
		\vspace{-0.2in}
        \caption{ICI-dominant regime: Velocity profile at $R = 120 \, \rm{m}$ obtained by the different methods \subref{fig_vp_ici_1} for the scenario in Fig.~\ref{fig_scenario_ici_dominant}, and \subref{fig_vp_ici_2} for a modified version of the scenario in Fig.~\ref{fig_scenario_ici_dominant}, where Target~3 and Target~4 move $10 \, \rm{m/s}$ faster.}  
        \label{fig_vp_ici}
        \vspace{-0.22in}
\end{figure}

We plot the velocity profiles obtained by the considered methods in Fig.~\ref{fig_vp_ici} for a single noise realization. As indicated in \eqref{eq_ici_exp} and Table~\ref{tab_parameters}, the proposed approach increases the maximum velocity by a factor of $N = 1024$, which allows detection of targets beyond the standard velocity limit $\vmax$. Hence, the proposed approach can resolve four targets and detect their true (unambiguous) velocities, while the FFT method can only detect two targets. This is observed also in Fig.~\ref{fig_vp_ici_2}, where four targets have different ambiguous velocities; the targets cannot be resolved by the FFT method due to poor velocity resolution. Through the ICI embracing capability of the proposed detector, all targets can be resolved.

\begin{figure}
        \begin{center}
        \subfigure[]{
			 \label{fig_pd_snr_ici}
			 \includegraphics[width=0.4\textwidth]{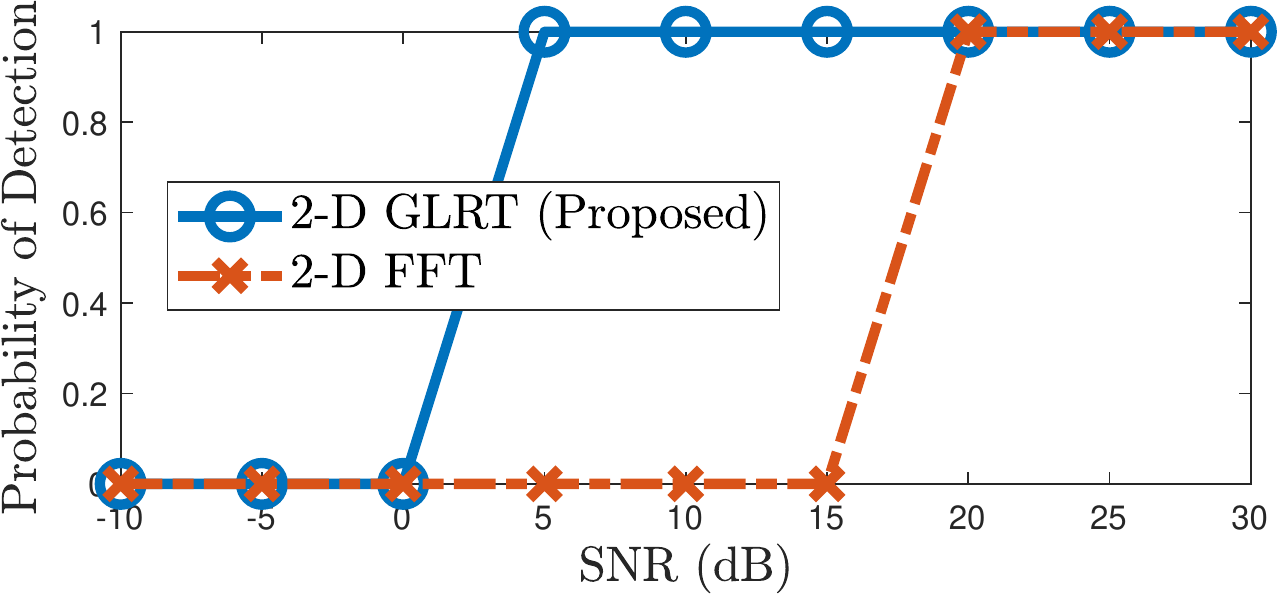}
		}
        \subfigure[]{
			 \label{fig_pfa_snr_ici}
			 \includegraphics[width=0.4\textwidth]{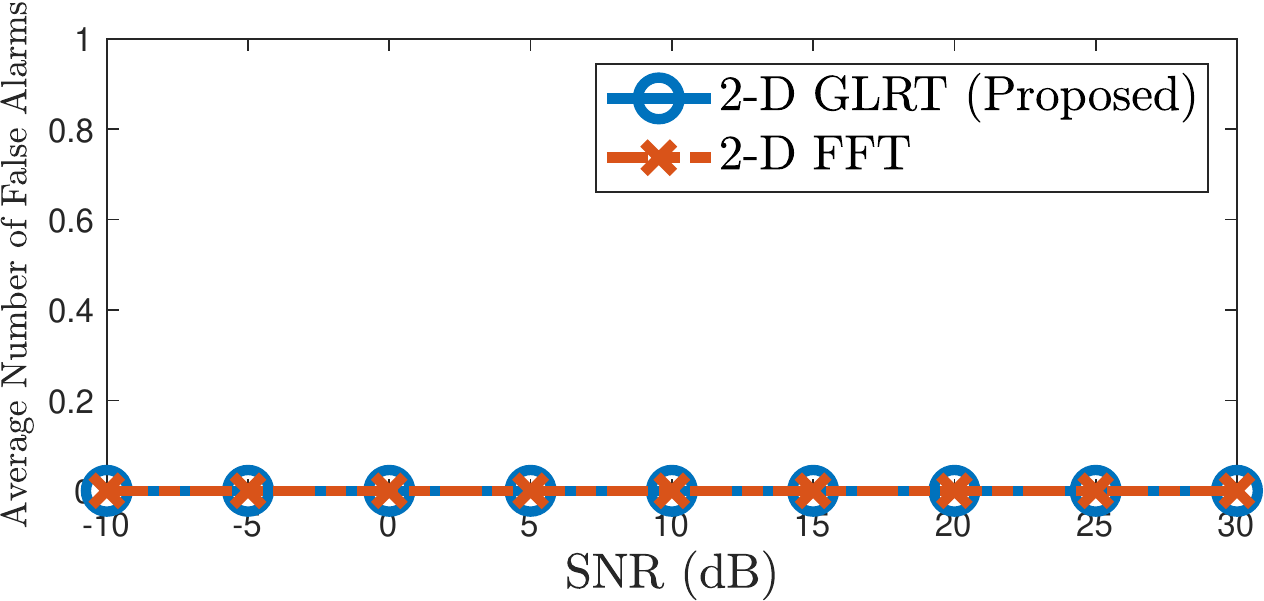}
		}
		
		\subfigure[]{
			 \label{fig_range_rmse_ici}
			 \includegraphics[width=0.4\textwidth]{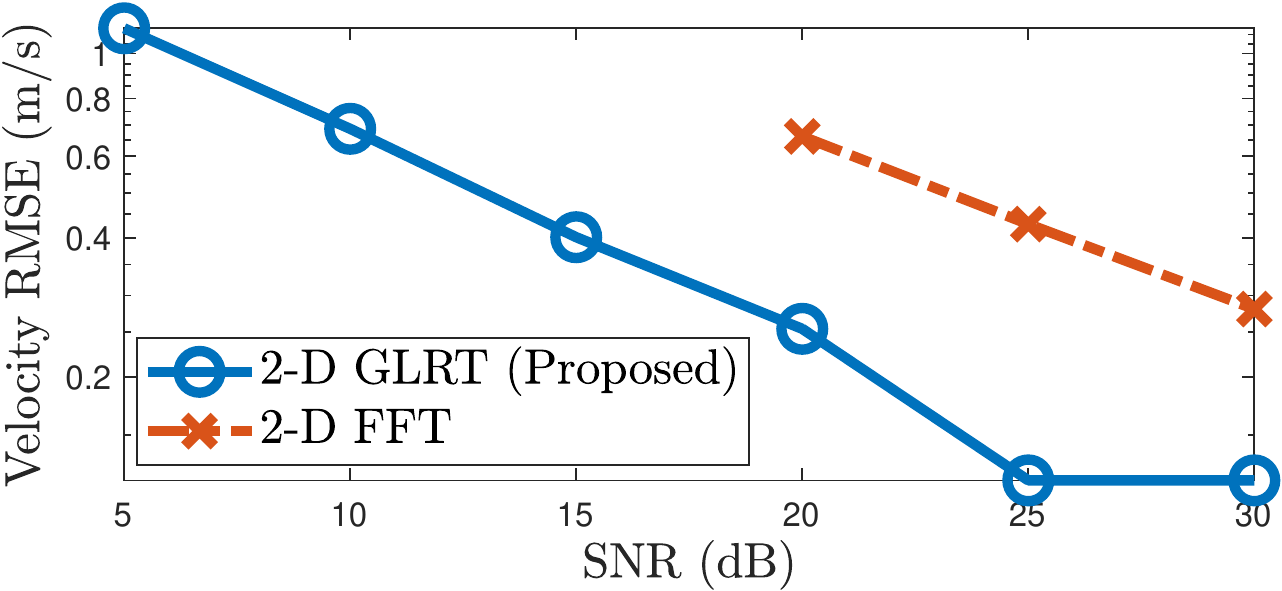}
		}

		\end{center}
		\vspace{-0.2in}
        \caption{Detection and estimation performance of the considered OTFS radar algorithms with respect to SNR in the ICI-dominant regime. \subref{fig_pd_snr_ici} Probability of detection, \subref{fig_pfa_snr_ici} average number of false alarms, and \subref{fig_range_rmse_ici} velocity RMSE.}  
        \label{fig_ici_det_est}
        \vspace{-0.2in}
\end{figure}

Fig.~\ref{fig_ici_det_est} shows the detection and estimation performance curves with respect to the SNR of the reference target, Target~2, averaged over $100$ realizations. Similar to the ISI-dominant case, the proposed approach achieves significant performance gains over the conventional FFT method in terms of both the probability of detection and velocity RMSE by explicitly accounting for the ICI effect in detector/estimator design.

\section{Concluding Remarks}\vspace{-0.05in}
In this paper, we have considered the problem of radar sensing with OTFS waveform and derived a low-complexity GLRT based detector/estimator by taking into account the ISI and ICI effects caused by the lack of guard intervals in time and frequency, respectively. The proposed radar processing approach not only compensates for the ISI and ICI effects, but also turns them into an advantage by significantly increasing maximum detectable range and velocity. Simulation results have verified the ISI and ICI embracing capability of the proposed radar receiver and demonstrated improved detection and estimation performance with respect to the standard FFT method employed in OFDM radar.

\section*{Acknowledgments}
\footnotesize{This work is supported by the EU Marie Curie Actions Individual Fellowship project OTFS-RADCOM, grant nr.~888913, Vinnova grant 2018-01929 and the European Commission through the H2020 project Hexa-X (Grant Agreement no.~101015956). The work of A. Alvarado is supported by the Dutch Technology Foundation TTW, which is part of the Netherlands Organisation for Scientific Research (NWO), under the project Integrated Cooperative Automated Vehicles (i-CAVE).}
\bibliographystyle{IEEEtran}
\bibliography{main}

\vspace{-0.1in}

\end{document}

%% file: commands.tex
\newcommand{\norm}[1]{\left\lVert#1\right\rVert}

\newcommand{\thnew}[1]{ {#1^{\rm{th} } } }

\newcommand{\Tcp}{ T_{\rm{cp}} }

\newcommand{\FF}{ \mathbf{F} }

\newcommand{\deltaf}{ \Delta f }
\newcommand{\fc}{ f_c }

\newcommand{\scp}{s_{\rm{CP}}}
\newcommand{\ycp}{y_{\rm{CP}}}

\newcommand{\gtx}{g_{\rm{tx}}}

\newcommand{\etatilde}{\widetilde{\eta}}

\newcommand{\pfa}{P_{\rm{fa}}}

\newcommand{\snr}{{\rm{SNR}}}

\newcommand{\xnm}{ x_{n,m} }

\newcommand{\deltar}{ \Delta R }
\newcommand{\deltav}{ \Delta v }
\newcommand{\Rmax}{ R_{\rm{max}} }
\newcommand{\vmax}{ v_{\rm{max}} }

\newcommand{\taumaxisi}{ \tau^{\rm{ISI}}_{\rm{max}} }
\newcommand{\rmaxisi}{ R^{\rm{ISI}}_{\rm{max}} }
\newcommand{\numaxici}{ \nu^{\rm{ICI}}_{\rm{max}} }
\newcommand{\vmaxici}{ v^{\rm{ICI}}_{\rm{max}} }

\newcommand{\taumax}{ \tau_{\rm{max}} }

\newcommand{\numax}{ \nu_{\rm{max}} }

\newcommand{\tauk}{ \tau_k }

\newcommand{\vk}{ v_k }
\newcommand{\nuk}{ \nu_k }

\newcommand{\mtCN}{{\mathcal{CN}}}

\newcommand{\diag}[1]{ {\rm{diag}}\left(#1\right)  }

\newcommand{\Imatrix}{{ \boldsymbol{\mathrm{I}} }}

\newcommand{\cc}{ \mathbf{c} }

\newcommand{\bb}{ \mathbf{b} }

\newcommand{\boldzero}{{ {\boldsymbol{0}} }}

\newcommand{\boldX}{ \mathbf{X} }
\newcommand{\boldXdd}{ \mathbf{X}^{\rm{DD}} }

\newcommand{\boldC}{ \mathbf{C} }
\newcommand{\boldB}{ \mathbf{B} }

\newcommand{\yy}{ \mathbf{y} }

\newcommand{\ww}{ \mathbf{w} }

\newcommand{\sss}{ \mathbf{s} }

\newcommand{\transpose}[1]{ {#1}^{T} }

\newcommand{\complexset}[2]{ \mathbb{C}^{#1 \times #2}  }